\documentclass[journal]{IEEEtran}
\usepackage{algorithm}
\usepackage[justification=centering]{caption}
\usepackage{algpseudocode}
\usepackage{fancyhdr}
\usepackage{textcomp}
\usepackage{amsmath}
\usepackage{subcaption}
\usepackage{amssymb,amsfonts}
\usepackage{graphicx}
\usepackage{textcomp}
\usepackage{xcolor}
\usepackage{multicol}
\usepackage{multirow}
\usepackage{blindtext}
\usepackage{adjustbox}
\usepackage[backend=bibtex,style=ieee]{biblatex}
\algnewcommand{\Initialize}[1]{%
  \State \textbf{Initialize:}
\parbox[t]{.8\linewidth}{\raggedright #1}
}
\algnewcommand{\Goto}{\textbf{go to}}%

\def\BibTeX{{\rm B\kern-.05em{\sc i\kern-.025em b}\kern-.08em
    T\kern-.1667em\lower.7ex\hbox{E}\kern-.125emX}}

\usepackage{graphicx}

\begin{document}
%
\title{Cooperation and Personalization on a Seesaw: Choice-based FL for Safe Cooperation in Wireless Networks}
%
%
%

    
    
    
\author{Han~Zhang,
        Medhat~Elsayed,
        Majid~Bavand,
        Raimundas~Gaigalas,
        Yigit~Ozcan,
        and~Melike~Erol-Kantarci,~\IEEEmembership{Senior Member, IEEE}
\thanks{Han Zhang and Melike Erol-Kantarci are with the School of Electrical Engineering and Computer Science, University of Ottawa, Ottawa, ON K1N 6N5, Canada (e-mail: hzhan363@uottawa.ca; melike.erolkantarci@uottawa.ca).}
\thanks{Medhat Elsayed, Majid Bavand, Raimundas Gaigalas and Yigit Ozcan are with the Ericsson, Ottawa, K2K 2V6, Canada(e-mail:
medhat.elsayed@ericsson.com; majid.bavand@ericsson.com; raimundas.
gaigalas@ericsson.com; yigit.ozcan@ericsson.com)
}}


\markboth{ }%
{Shell \MakeLowercase{\textit{et al.}}: Bare Demo of IEEEtran.cls for IEEE Journals}

\maketitle
\thispagestyle{fancy}         
\fancyhead{}                  
\lhead{This paper has been accepted by IEEE Wireless Communication Magazine.}

\begin{abstract}
Federated learning (FL) is an innovative distributed artificial intelligence (AI) technique. It has been used for interdisciplinary studies in different fields such as healthcare, marketing and finance. However the application of FL in wireless networks is still in its infancy. In this work, we first overview benefits and concerns when applying FL to wireless networks. Next, we provide a new perspective on existing personalized FL frameworks by analyzing the relationship between cooperation and personalization in these frameworks. Additionally, we discuss the possibility of tuning the cooperation level with a choice-based approach. Our choice-based FL approach is a flexible and safe FL framework that allows participants to lower the level of cooperation when they feel unsafe or unable to benefit from the cooperation. In this way, the choice-based FL framework aims to address the safety and fairness concerns in FL and protect participants from malicious attacks. 
\end{abstract}

%
\IEEEpeerreviewmaketitle

\section{Introduction}
The growing scale and heterogeneity of wireless networks have boosted the applications of data-driven artificial intelligence (AI) algorithms in wireless networks. Federated learning (FL) is an innovative distributed AI technique that allows participants to train models cooperatively without sharing data. It has been rapidly promoted in recent years by the research community with the benefits of privacy preservation and the utilization of on-device intelligence \cite{zhang2023device}. 
However, the practical application of FL in wireless networks is still in its infancy. It has not yet gained widespread acceptance in the industry, mainly due to the security concerns and inherent problems of the cooperative mechanism. The distributed and open system architecture introduces new threats and makes FL models more vulnerable to poisoning attacks compared with centralized machine learning models \cite{zhang2023distributed}.

Several existing studies explore how to defend against vulnerabilities and enhance the security of FL models. Some commonly used defense schemes include anomaly detection, trusted execution authentication, and robust model aggregation. Nevertheless, these state-of-the-art defense techniques have adopted a sporadic approach, incorporating add-on techniques into the FL framework in a fix-and-patch manner. In addition, the attackers are becoming intelligent and can learn to adapt to existing defense schemes \cite{li2022learning}. This makes defenses more difficult, especially without knowing the type and principle of attacks. Therefore, there is still great potential to improve and promote the security level of FL in wireless communications. 

\begin{figure*}[t]
\centering
\includegraphics[width=6.8in]{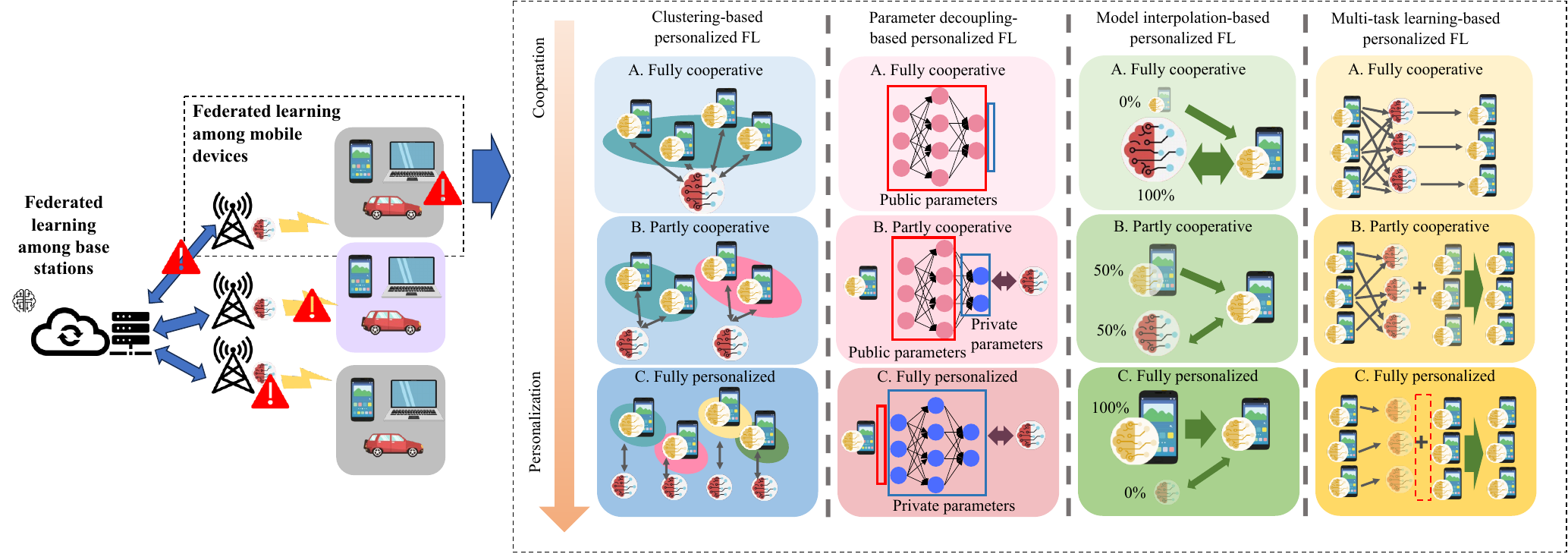}
\caption{The examples of tunable cooperation in FL and their applications to FL-based wireless networks. The left part shows the applications and vulnerabilities of FL in wireless networks. The right part shows examples of tunable cooperation in existing personalized FL techniques. By changing the parameters related to the cooperation degree, given personalized FL techniques will convert between fully cooperative FL and fully personalized independent learning.}
\label{fig1}
\end{figure*}

Personalized FL schemes have been proposed by academia as effective solutions to the problem of local data and problem-setting heterogeneity in FL. A thorough survey is presented in \cite{tan2022towards} on recent advances in personalized FL and gives a taxonomy of personalized FL techniques in terms of personalization strategies. However, most existing studies focus on describing detailed implementations of different personalized FL approaches, with limited exploration of the commonalities in these designs. 

After an exhaustive investigation of the personalized FL, we propose a new understanding of the essence of existing personalized FL frameworks. To the best of our knowledge, this is the first time that the tunable cooperation concept in personalized FL is proposed and discussed. In our view, the traditional FL schemes create a fixed cooperative paradigm for multiple independent participants. Participants must be fully committed to the cooperation and share the same cooperation output. On this basis, personalized FL methods create more flexible cooperative paradigms by introducing parameters that regulate the closeness degree of cooperation between participants. Therefore, we conclude that cooperation and personalization in FL are two ends of a seesaw. By controlling the parameters representing the degree of personalization and cooperation, it is possible to control which side of the seesaw is higher according to the problem and environment settings, thus realizing tunable cooperation in FL. 

Motivated by the above ideas, we introduce a flexible choice-based FL framework that dynamically adjusts the degree of cooperation and personalization of participants during the use of FL. It is also a more equitable and safer framework that allows participants to not cooperate when they feel unsafe or unable to benefit from cooperation. Note that, our choice-based FL can not only be used to solve the heterogeneity problem of traditional FL but also can be applied to noise and security concerns. When the system is under attack, participants will be more inclined to personalized training to avoid interference from attacks and noise. Through these structural changes, FL is given a higher degree of flexibility and can be more widely used in wireless networks. 

The contributions of this work are three-fold: First, we present an up-to-date survey of FL in wireless networks and illustrate the benefits and concerns of FL. Second, we provide a new perspective on the commonalities of existing personalized FL algorithms and propose the concept of tunable cooperation. Based on this concept, we introduce a flexible choice-based FL framework that solves major concerns of FL, including heterogeneity, fairness, and poisoning attacks. Third, we provide a case study of implementing a choice-based FL framework with clustering and knowledge distillation scheme under a cell sleep control scenario. By testing the framework with heterogeneous data and intelligent attacks, we prove the superiority of choice-based FL in terms of robustness and security.

The remainder of this work is organized as follows. Section II explains the benefits and vulnerabilities of FL in wireless networks. Section III presents the tunable cooperation concept in personalized FL. Section IV introduces the choice-based FL framework and shows how it solves the vulnerabilities of FL. Section V demonstrates our case study and Section VI concludes the paper.

\section{Benefits and Concerns of Federated Learning in Wireless Networks}
\label{s2}

The left part of Fig. \ref{fig1} shows the possible applications and vulnerabilities of FL in wireless networks. It is worth noting that FL can be used for two different objectives. In the first case, FL training intends to generate a good global model with distributed data. In the second case, FL training intends to provide good local models for clients since the clients are the actual users of the trained model. Applications in wireless communication tend to be the latter. In this paper, we overview both types of approaches but our proposed scheme focuses on the latter case.

In the following, we give a more in-depth examination of the benefits and concerns of applying FL in wireless communications.

\subsection{Benefits of Federated Learning}
The benefits of FL are summarized below:
\begin{itemize}
    \item Privacy: Different from centralized learning methods, FL participants do not upload raw data to the global servers. They only upload model parameters, thus avoiding data leakage.
    \item Scalability: As a distributed learning framework, FL can handle large-scale data by leveraging on-device computational and memory resources.
    \item Communication efficient and time-saving: FL trains models locally and skips the step of collecting and aggregating data from diverse sources. This cooperative pattern improves communication efficiency by avoiding data transfer and saves model training time.
    \item Robustness: The adversarial robustness in FL can be propagated from rich-resource participants to those with poor resources, thus improving the general performance of the system \cite{hong2023federated}.
    \item Diversity and inclusiveness: FL allows the use of data that cannot be collected directly as part of the training dataset. It helps reduce bias and develop a more representative and inclusive model \cite{kairouz2021advances}.
\end{itemize}

\subsection{Concerns about Federated Learning}

Despite the benefits mentioned above, FL algorithms also suffer from several concerns. Some main concerns are listed as follows: 
\begin{itemize}
    \item Fairness among participants: Participants involved in FL usually carry varying amounts of data and computational power. Without well-developed incentives, it will be unfair to the resourceful participants.
    \item Heterogeneity: The heterogeneity of participants in FL includes the heterogeneity of data, the heterogeneity of model structure, and the heterogeneity of training methods. 
    \item Trade-off between the security level and the system performance: There is an intrinsic conflict between the system security level and the performance of FL. Some defense methods affect model performance when they attempt to improve the security level.
\end{itemize}

The potential vulnerabilities that arise from the distributed structure of FL are one of the main concerns of FL. As shown in the left part of Fig. 1, the vulnerabilities come from both inside (FL servers and clients) and outside (external communication links). Some common types of attacks and the consequences of the attacks are listed as follows:
\begin{itemize}
    \item Free-riders: Free-riders refer to ill-intentioned participants in FL who aim to receive the aggregated global model without contributing any useful local updates. Such attacks compromise the fairness of the cooperation and discourage benign participants.
    \item Poisoning attacks: Poisoning attack means that the local model or data of one or more participants is maliciously tampered with. The poisoning attacks will affect the model training and lead to degradation in network performance.
    \item Backdoor attacks: Backdoor attacks inject malicious data with specific trigger patterns. They are more stealthy than regular data poisoning attacks since they cause the model to misbehave only under trigger conditions and behave normally in other cases.
    \item Malicious servers: Malicious servers refer to attacks when the server of FL is malicious or compromised. These attacks tamper with the global model aggregation process and corrupt the global model. 
\end{itemize}

In the following sections, we will explore the essence of personalized FL and explain how to leverage it to mitigate the concerns of FL.
\section{Tunable Cooperation in Personalized Federated Learning}
In traditional FL algorithms such as FedAvg \cite{mcmahan2017communication}, participants will be treated the same. They are asked to devote all their efforts of local training to cooperation and share the same global model as the basis for local updates. That also means participants will fully trust the other participants and the global server. 

To solve the data heterogeneity challenge of traditional FL algorithms, personalized FL algorithms have been proposed to improve the performance on local data of each participant. This is achieved by designing personalized branches and customized local models for different participants. Some widely used personalized FL approaches include clustering, parameter decoupling, model interpolation, multi-task learning, knowledge distillation, meta-learning, and regularized local loss \cite{tan2022towards}. In the following, we provide a more macroscopic explanation of existing personalized FL algorithms, showing that the essence of personalized FL lies in the introduction of a tunable cooperation mechanism. We further show how such mechanisms can benefit FL not only in terms of heterogeneity but also in other aspects, such as security and fairness.

\begin{figure*}[t]
\centering
\includegraphics[width=6.8in]{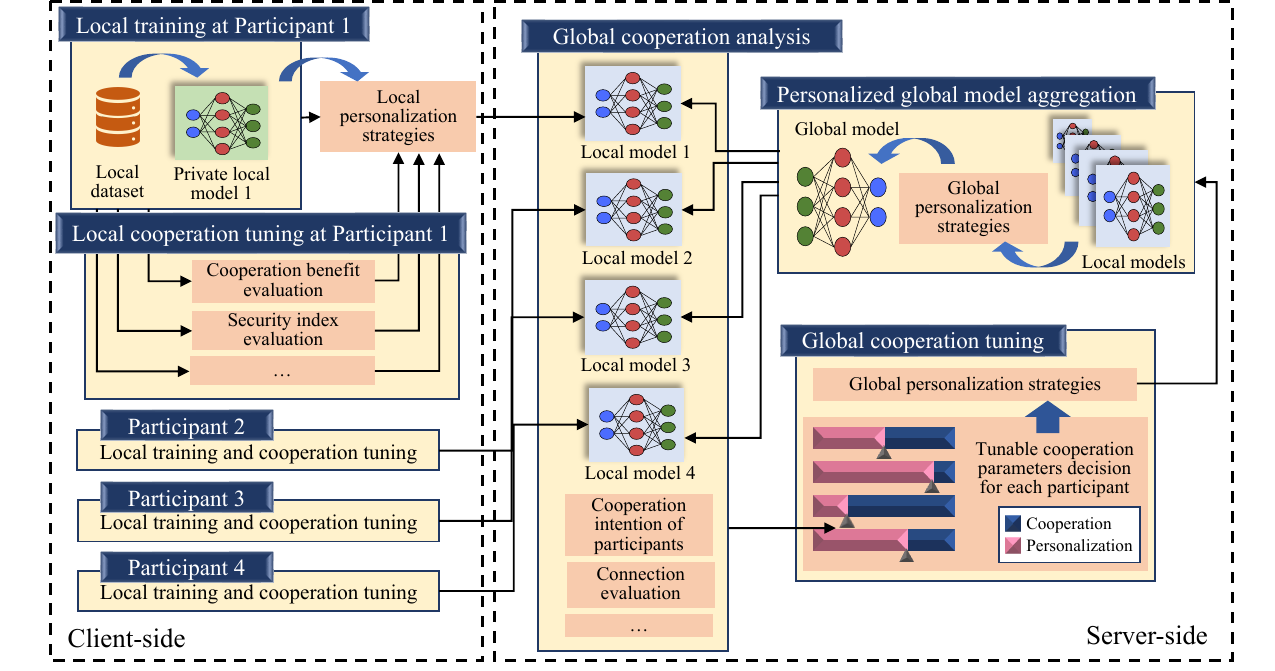}
\caption{The pipeline of the proposed choice-based FL.}
\label{fig3}
\end{figure*}

From a general point of view, personalized FL algorithms are an intermediate form between traditional fully cooperative FL and fully independent distributed training. The way to reach this intermediate form is to introduce one or more additional parameters to control the closeness of cooperation between participants, which is called tunable cooperation. The right part of Fig. \ref{fig1} gives examples of tunable cooperation under four different personalized FL algorithms. In each column, we give a common personalized FL framework and explain how the framework varies between the fully cooperative and fully personalized states under tunable cooperation. 
The detailed explanations are given as follows:
\begin{itemize}
    \item Clustering-based personalized FL: In clustering-based personalized FL, participants are firstly clustered and FL is performed independently inside each cluster. Personalized global models are developed for participants within different clusters. The clustering is usually performed according to the similarity of the local updates of clients to the global model, or the clustering strategy can be optimized by minimizing the global loss of the FL. In this framework, the parameter that controls the tunable cooperation between participants is seen as the number of clusters or the similarity threshold for generating new clusters. In an extreme case when there is only one cluster, the participants will work in a fully cooperative manner, and the clustering-based personalized FL will transform into a FedAvg framework. Meanwhile, if the cluster number is identical to the number of participants, each participant will be a cluster, and the framework will transform into a fully independent distributed learning.
    \item Parameter decoupling-based personalized FL: In parameter decoupling-based personalized FL, model parameters are decoupled as private and public local model parameters. Only public local model parameters will be shared with the global server for cooperation, while private local model parameters will be kept locally for personalization. In this framework, the parameter that controls the tunable cooperation between participants is seen as the ratio of private to public parameters. In an extreme case when all the parameters are public parameters, the participants will be fully cooperative, and the framework is equivalent to FedAvg. Instead, if all the parameters are private parameters, the participants will train their models fully independently and there is no cooperation between participants.
    \item Model interpolation-based personalized FL: In model interpolation-based personalized FL, local personalization is performed by using a mixture of global and local models for participants. Similarly, the framework converts to fully cooperative FL if the  proportion of global models in the mixture is extremely large. Instead, it will be converted to fully independent distributed learning if the proportion of global models in the mixture is extremely small.
    \item Multi-task learning-based personalized FL: Multi-task learning-based personalized FL regards each participant in FL as a task. Instead of using one single global model, each participant keeps a cloud model, which is a linear combination of all the local models. In this case, the closeness of cooperation is regulated by the generation of the cloud model. If all the participants have identical cloud models and each local model shares the same weight in the linear combination, the framework will turn into FedAvg. Meanwhile, if the cloud model of each participant is identical to its local model, the participants will be trained independently.
\end{itemize}

This paper, for the first time, collects personalized FL techniques under the umbrella of tunable FL showing that different aspects of personalization are in fact, a result of a parameter that tunes the degree of cooperation. Although different parameters are tuned, as seen in the above-introduced techniques, a general concept of tunable parameters has been missing in the literature. To this end, we group the tunability approaches and develop a more generic framework in the following sections.

\begin{table*}[]
\centering
\caption{Summary of personalized FL algorithms and how they defend against poisoning attacks from malicious participants.}
\label{table1}
\resizebox{2\columnwidth}{!}{%
\begin{tabular}{|c|c|cc|c|}
\hline
\multirow{2}{*}{\textbf{Algorithms}} &
  \multirow{2}{*}{\textbf{Parameter for cooperation tuning}} &
  \multicolumn{2}{c|}{\begin{tabular}[c]{@{}c@{}}\textbf{Personalization}\\ \textbf{is performed at}\end{tabular}} &
  \multirow{2}{*}{\textbf{Approach for mitigating poisoning attacks for local models}} \\ \cline{3-4}
 &
   &
  \multicolumn{1}{c|}{\begin{tabular}[c]{@{}c@{}}\textbf{Global} \\ \textbf{server}\end{tabular}} &
  \begin{tabular}[c]{@{}c@{}}\textbf{Local} \\ \textbf{clients}\end{tabular} &
   \\ \hline
Clustering \cite{ghosh2020efficient} &
  Number of clusters &
  \multicolumn{1}{c|}{\checkmark} &
   &
  \begin{tabular}[c]{@{}c@{}}Perform fine-grained clustering by increasing clustering \\ number to group the attackers into one category\end{tabular} \\ \hline
Parameter decoupling \cite{arivazhagan2019federated} &
  \begin{tabular}[c]{@{}c@{}}Number of globally shared \\ parameters or layers\end{tabular} &
  \multicolumn{1}{c|}{\checkmark} &
  \checkmark &
  \begin{tabular}[c]{@{}c@{}}Decrease the globally shared parameters and keep \\ poisoned parameters locally\end{tabular} \\ \hline
Model interpolation \cite{hanzely2020federated} &
  \begin{tabular}[c]{@{}c@{}}The proportion of the global \\ model and local models \\ in the mixture\end{tabular} &
  \multicolumn{1}{c|}{\checkmark} &
  \checkmark &
  \begin{tabular}[c]{@{}c@{}}Decrease the proportion of the global model and make \\ benign local models less affected by attacks\end{tabular} \\ \hline
Multi-task learning \cite{smith2017federated} &
  \begin{tabular}[c]{@{}c@{}}The coefficients of the linear \\ combination that makes \\ up the cloud model\end{tabular} &
  \multicolumn{1}{c|}{} &
  \checkmark &
  \begin{tabular}[c]{@{}c@{}}Build cloud models for benign participants only \\ with other benign participants based on \\ model similarity measurements\end{tabular} \\ \hline
Transfer learning \cite{chen2020fedhealth} &
  \begin{tabular}[c]{@{}c@{}}Iterations of local \\ model adaption\end{tabular} &
  \multicolumn{1}{c|}{} &
  \checkmark &
  \begin{tabular}[c]{@{}c@{}}Perform more iterations of local adaption to correct \\ attacks from the global model\end{tabular} \\ \hline
Knowledge distillation \cite{zhu2021data} &
  \begin{tabular}[c]{@{}c@{}}The threshold that \\ controls the direction \\ of knowledge flow\end{tabular} &
  \multicolumn{1}{c|}{\checkmark} &
  \checkmark &
  \begin{tabular}[c]{@{}c@{}}Only distill useful knowledge and filter filter attacks\\ during knowledge distillation\end{tabular} \\ \hline
Reward shaping \cite{hu2021reward} &
  \begin{tabular}[c]{@{}c@{}}The proportion of \\ global model-based reward\end{tabular} &
  \multicolumn{1}{c|}{} &
  \checkmark &
  \begin{tabular}[c]{@{}c@{}}Decrease the portion of global model-based reward\\ to make benign local models less affected by attacks\end{tabular} \\ \hline
\end{tabular}%
}
\end{table*}

The above conclusion of the personalized FL methodology gives an insight into the feasibility of making structural changes to FL. To balance between cooperation and personalization, the cooperation degree of participants in FL should not be fixed, but dynamically changed according to the conditions of the moment. In other words, cooperation and personalization should be the choice of participants. By analyzing environmental information and their own needs, participants in FL make rational choices about the cooperation degree to guarantee the optimality of model performance. In the next section, we introduce the choice-based FL framework to implement automated dynamic cooperation tuning in FL.
\section{Choice-based Tunable Federated Learning Framework}

Based on the idea of tunable cooperation, we designed a choice-based FL framework to allow participants and the global server to choose the level of cooperation in the system. Fig. \ref{fig3} illustrates the pipeline of the choice-based FL framework. This framework applies to most personalized FL methods. Before applying it to a specific personalized FL method, we first need to determine where the method is performed and which parameter is used for cooperation tuning. Such information for some common personalization FL methods is summarized in Table \ref{table1}. The personalized FL methods can be operated in two possible locations. Some of the methods, such as clustering, run on the global server side, while other methods, such as transfer learning, run on the client side. Accordingly, in the choice-based FL framework, we set up both the local cooperation tuning module and the global cooperation tuning module.

In our proposed choice-based FL framework, cooperation analysis is first conducted to assess how closely the participants should cooperate. The specific method of performing analysis will be introduced in the subsequent paragraphs. We define the independently distributed learning framework as having a 0\% level of cooperation and the FedAvg framework as having a 100\% level of cooperation. By controlling the parameters related to cooperation tuning in the personalized FL, the system can be made to shift between the independent distributed learning state and the fully cooperative FedAvg state. In this way, participants in the FL system can benefit from cooperation while retaining personalization.

The local cooperation analysis is performed by the participants to assess how closely they are willing to cooperate. The analysis can be based on considerations including how reliable the global model is and how much the participants benefit from the cooperation. There are two ways to assess these indicators. First, the participants can compare the received global model parameters with local model parameters. If the global model parameters have significantly deviated from the local ones, then the global model is likely to be unreliable. In this case, participants will be more inclined to train the model independently rather than engage in collaboration. Second, the participants can validate the received global model on its local dataset. Specifically, they can feed the local training data into the global model to see if they can predict the correct results. Or they can input the same data to the global model and the local model to compare the similarity of the output distributions. If the global model validates poorly on the local dataset or the output distributions are very different, the participant is likely not to benefit from cooperation. Therefore, the participant may want to reduce cooperation and enhance personalization to ensure the local model performance. Instead, if the global model validates well on the local dataset or the output distributions are very close, participants will benefit from the cooperation and they will increase the cooperation level. Therefore, we can quantify the accuracy of the local validation or the similarity of the distributions as a value between 0 and 1, and use this value as the cooperation level expected by the participant. After deciding on the cooperation level, local personalization strategies are developed by adjusting the parameters for cooperation tuning. Then local models are sent to the global server for global aggregation.

For personalized FL algorithms that are performed at the global server, the global cooperation analysis and global cooperation tuning are performed during the global aggregation. During the global cooperation analysis, the cooperation level is determined by the cooperation intention indicated by the participants, the correlation between the local models, and other plausible considerations. Similarly, the global server can convert these indicators into a value between 0 and 1, and use this value as the cooperation level. After that, the global server will develop global personalization strategies by changing the parameters for cooperation tuning accordingly during the global aggregation. The personalization strategies help to develop personalized local models, which will then be sent back to local servers for model updates in the next iteration.

The advantages of the choice-based FL framework are evident. It can mitigate most of the concerns of FL we have mentioned in the previous sections. First, this is a generic framework for both homogeneous and heterogeneous participants. The embedding of personalization strategies addresses heterogeneity concerns of FL. Second, the choice-based FL framework provides a fair platform for participants since they can adjust their contribution to the cooperation according to the benefits they receive. If the participants do not benefit from cooperation, they can reduce the cooperation level to focus more on local training. This is usually more in line with the expectations of participants for the cooperative training mechanism. In addition, the proposed framework also protects benign participants from malicious attacks. Briefly, when a malicious attacker wants to carry out an attack, the global server and the benign participants will find out from the cooperation analysis that the current situation is not suitable for close cooperation. Therefore, they will reduce the cooperation level and limit the impact of the attacks. Table \ref{table1} also explains why the common personalized FL methods can defend against malicious attacks by decreasing the level of cooperation.

\section{Case Study on Cell Sleep Control}

\begin{figure}[t]
\centering
\includegraphics[width=3.5in]{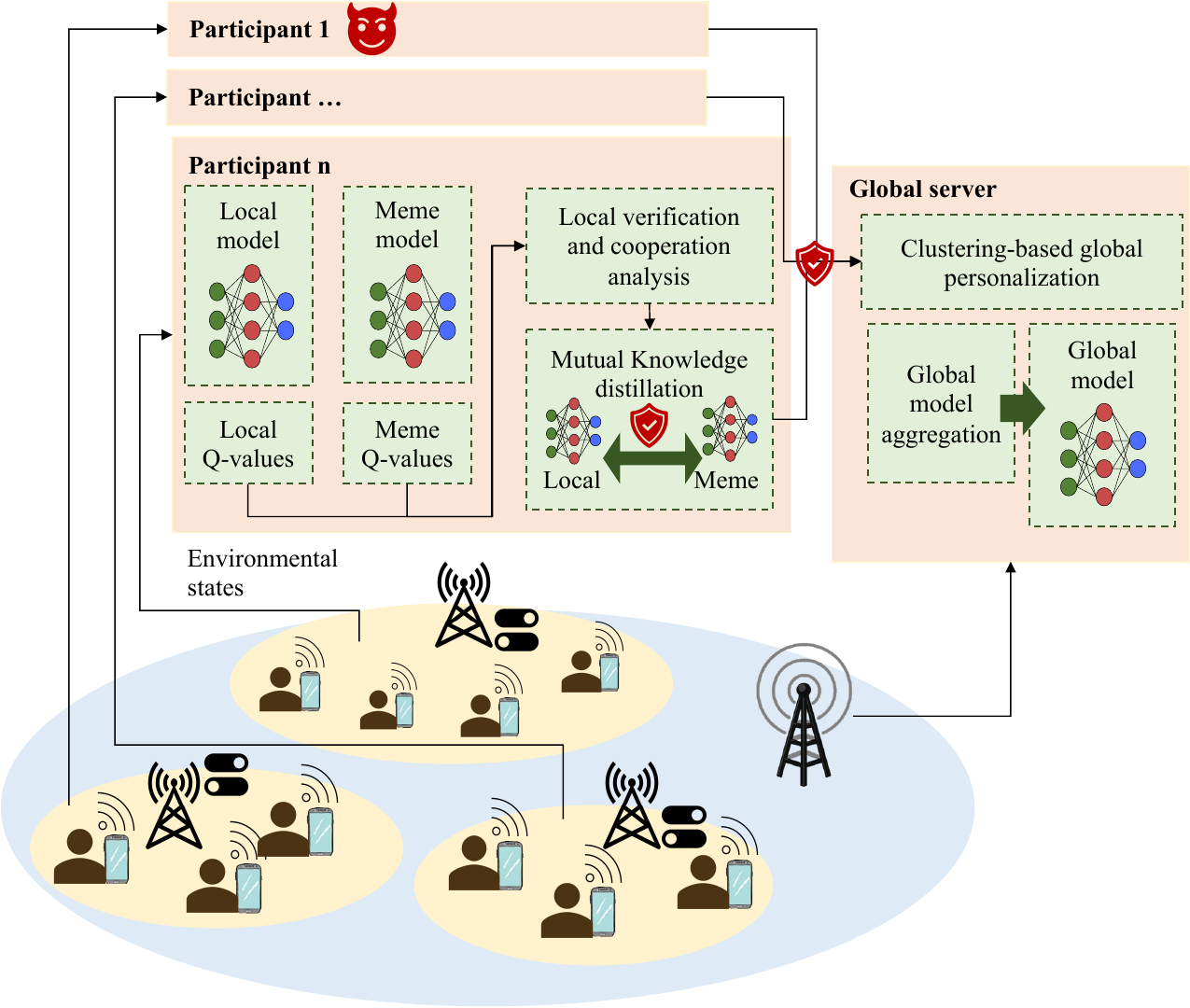}
\caption{The system model of knowledge distillation-enabled FL-based cell sleep control.}
\label{fig7}
\end{figure}

This section presents a case study on a knowledge distillation-enabled choice-based FL framework for the cell sleep control application, following our prior work in \cite{zhang2023distributed}. Fig. \ref{fig7} shows the system model and algorithm design of the given application. We assume a heterogeneous wireless network with one macro base station (MBS) and several small base stations (SBSs). Each SBS holds a local Deep Q Network (DQN) model to perform dynamic cell sleep control. At the same time, the MBS always stays active to ensure coverage, and it also serves as a global server to aggregate local models from SBSs. The SBS chooses from three different sleep modes: active, sleep, and deep sleep. The objective of sleep control for SBS is to achieve high throughput and high energy efficiency. 

Instead of directly aggregating local models and distributing global models, mutual knowledge distillation is performed between the global and local models by setting up an intermediate memory model called the meme model. In each round, the parameters of the meme model, rather than the local model, are submitted to the global server for global model aggregation. Then, the global model parameters are sent back to the meme models. To dynamically change the cooperation level between participants, we control the direction of knowledge flow between the meme model and the local model and regulate the proportion of knowledge learned by the local model from the local data and the global model. Beyond this, a clustering-based global personalization method is performed to further remove the exceptional participants in global aggregation. 

Instead of using an off-the-shelf dataset, we train the DQN model online by interacting with the wireless communication environment. To evaluate whether the proposed choice-based FL framework can protect the participants from malicious attacks, we first simulated a near-realistic urban network environment with Python and built the channel model and communication model according to the 3GPP standard. Next, we generated traffic data according to the 24-hour typical residential area traffic pattern \cite{6056691} and modeled the data transmission between the SBS and the UE. In this process, the Markov decision process (MDP) data was generated and collected at each SBS. Compared with simulations with constant or random traffic patterns, simulations with the 24-hour traffic pattern are more realistic.

\begin{figure*}[ht]
\centering
 \begin{subfigure}[b]{0.45\textwidth}
     \centering
     \includegraphics[width=\textwidth]{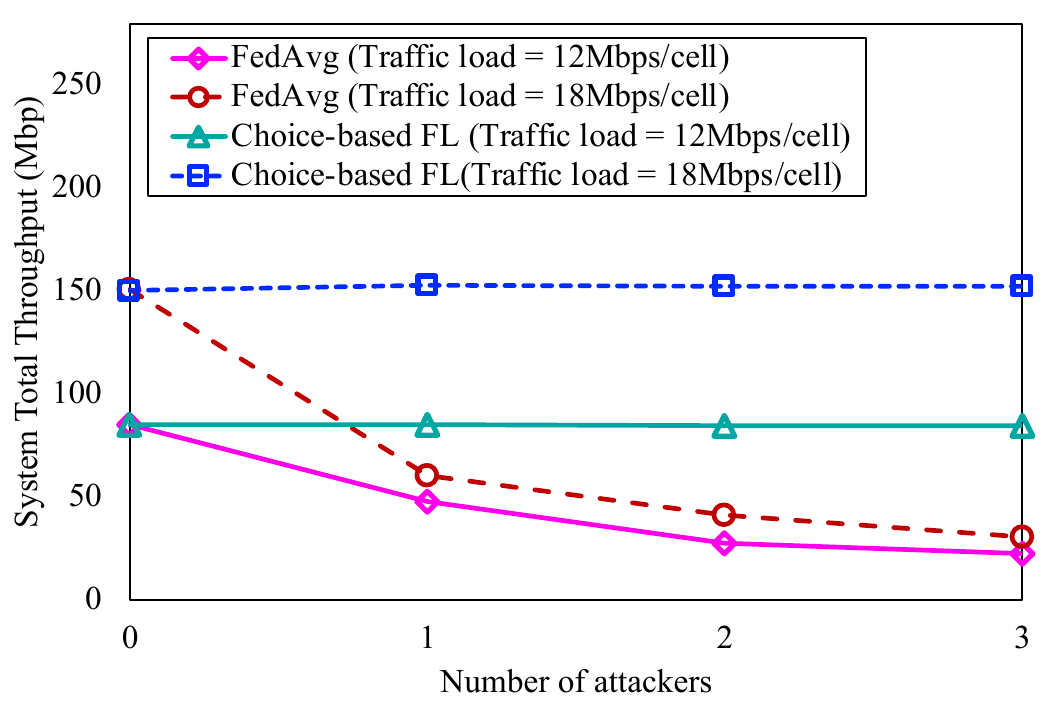}
     \captionsetup{font={small}}
     \caption{System total throughput under model poisoning attacks.}
     \label{fig5-1}
 \end{subfigure}
  \hfill
 \begin{subfigure}[b]{0.45\textwidth}
     \centering
     \includegraphics[width=\textwidth]{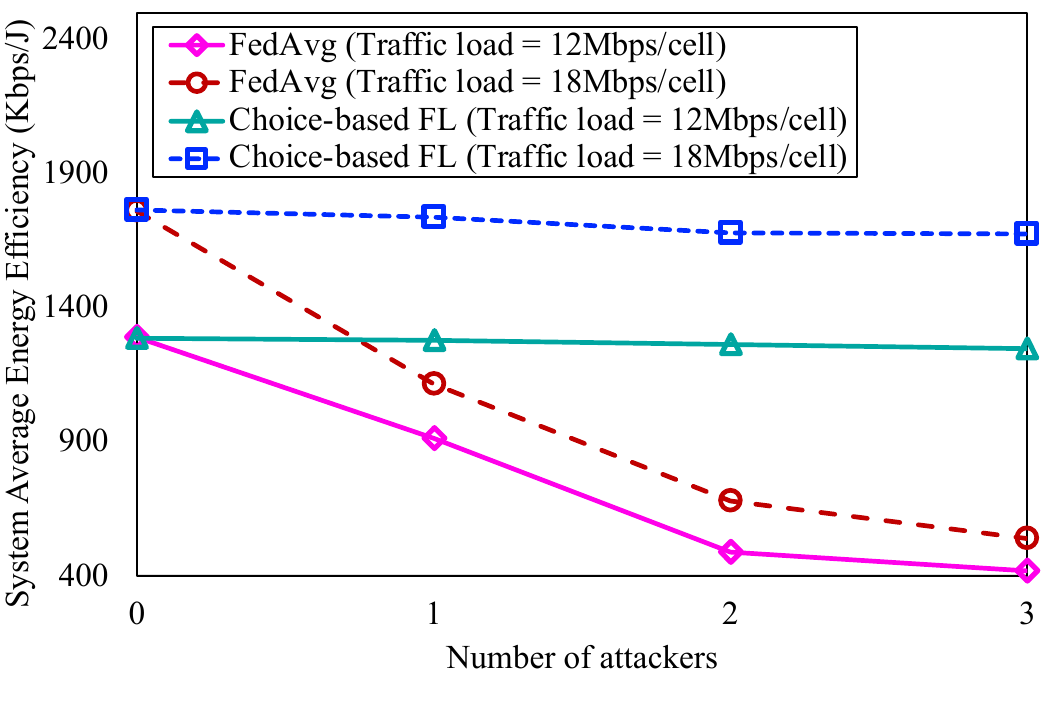}
     \caption{System average energy efficiency under model poisoning attacks.}
     \label{fig5-2}
 \end{subfigure}
  \begin{subfigure}[b]{0.45\textwidth}
     \centering
     \includegraphics[width=\textwidth]{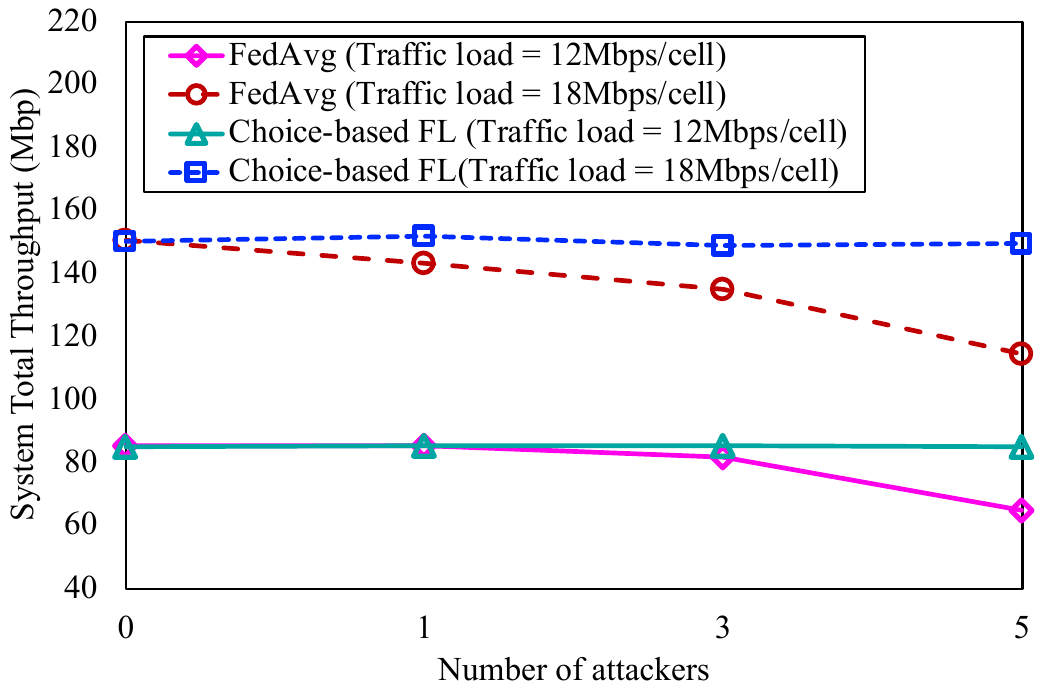}
     \captionsetup{font={small}}
     \caption{System total throughput under data poisoning attacks.}
     \label{fig5-3}
 \end{subfigure}
  \hfill
 \begin{subfigure}[b]{0.45\textwidth}
     \centering
     \includegraphics[width=\textwidth]{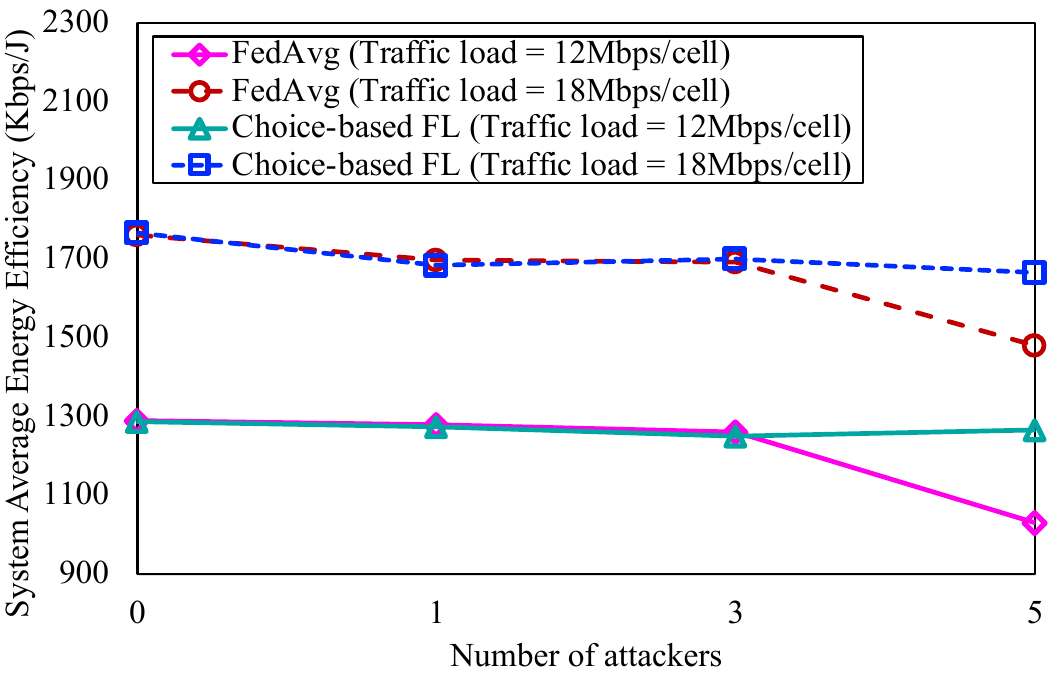}
     \caption{System average energy efficiency under data poisoning attacks.}
     \label{fig5-4}
 \end{subfigure}
\caption{System total throughput, average energy efficiency and cooperation level under different numbers of attackers.}
\label{fig5}
\end{figure*}

\begin{figure}[h]
\centering
\includegraphics[width=3.5in]{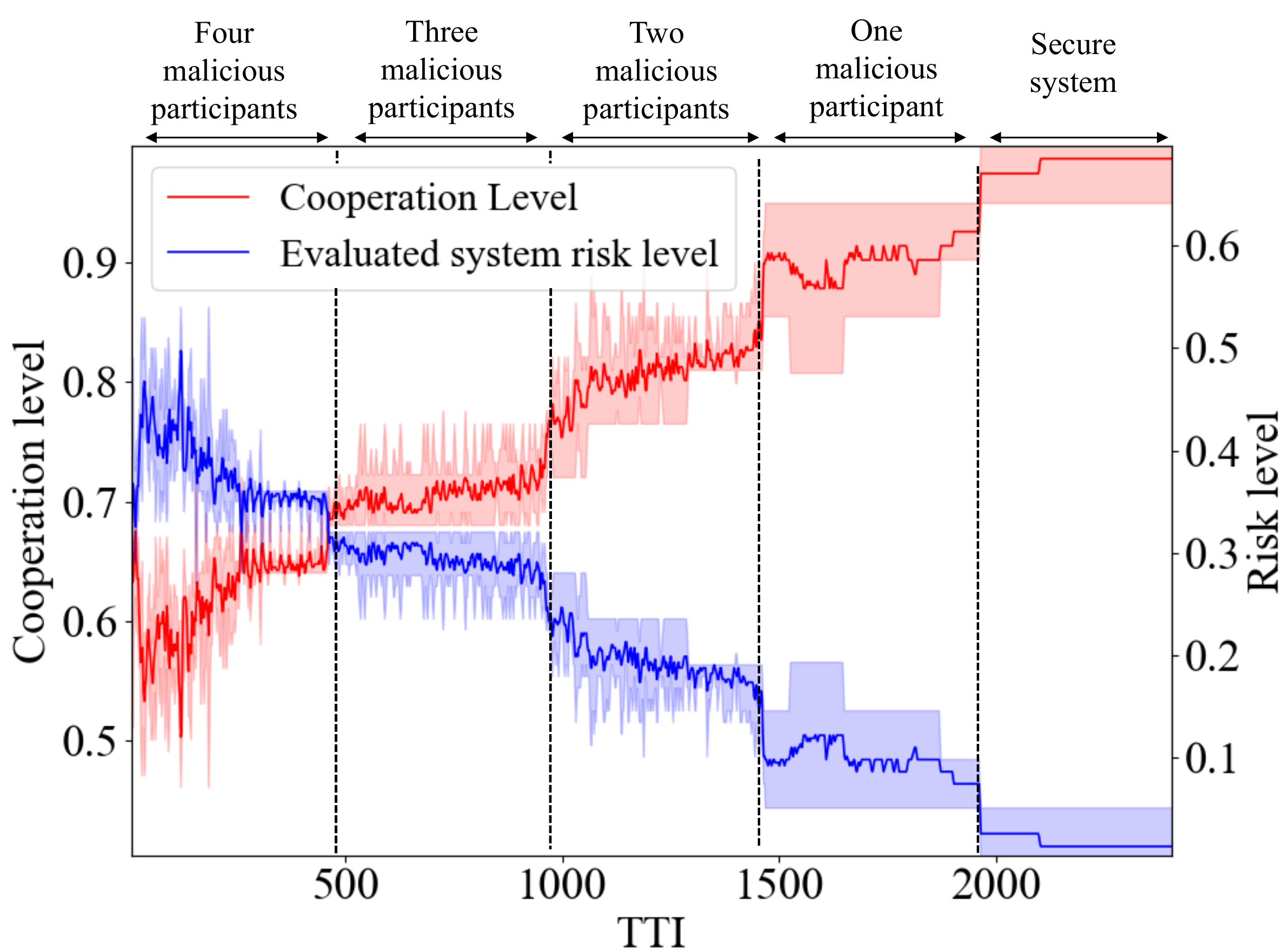}
\caption{The cooperation level and system risk level under different numbers of attackers.}
\label{fig6}
\end{figure}

To ensure that the MDP data collected by SBSs is non-IID and sufficiently challenging for FL, we set different numbers of UEs and different traffic loads for each SBS. In addition, we further checked the distribution of the local model parameters after local training to verify that they are also non-IID.

The simulation includes two parts. The first part is to demonstrate that our proposed framework can mitigate the impact of malicious attacks on FL-based wireless communication control problems. In this part, two types of malicious attacks are considered during the simulation, the data poisoning attack and the model poisoning attack. The model poisoning attack directly generates malicious local model parameters by changing the direction of the gradient descent, while the data poisoning attack injects malicious data into the training dataset of the compromised SBSs. We conducted several experiments, each using a fixed number of attackers, and measured the system performance after the model converged. The simulation results are shown in Fig. \ref{fig5}. The second part of the simulation aims to validate the tunable cooperation concept mentioned in Section III and Section IV. In this part, we perform a model poisoning attack and increase the attack intensity by gradually increasing the number of attackers. Then we observe the variation of the system risk level and cooperation level during the whole simulation process. The simulation results are shown in Fig. \ref{fig6}. In the following, we give a more detailed description and analysis of the results.

Fig. \ref{fig5-1} and Fig. \ref{fig5-2} show the system throughput and energy efficiency under model poisoning attacks. During the simulation, we observe the system total throughput and the energy efficiency with different numbers of attacked SBSs. We compare the results of using the FedAvg framework and the choice-based FL framework. It is observed that with FedAvg, participants work in a fully cooperative fashion. While high throughput and energy efficiency can be achieved in a secure environment, this framework also makes the participants more vulnerable to attack. The throughput and energy efficiency both decline rapidly as the number of attacked base stations increases. 

In contrast, the throughput and energy efficiency of the choice-based FL remain unchanged when the number of attackers increases. This means that the system performance of choice-based FL is consistent with that in the secure system (indicated by the number of attackers as 0). This is because choice-based FL provides a strategy to dynamically tune the degree of cooperation between participants, leading to a more robust system performance. Without any attack, the participants will detect a secure environment and automatically enhance cooperation. At this point, choice-based FL achieves close system throughput and energy efficiency when fully cooperated. With attacks, the participants will weaken the cooperation, focusing more on personalized training to protect their local models from attacks. In this way, malicious attacks will not be delivered to benign participants through global aggregation and cooperation. As can be observed from the results, the system performance can maintain a comparable level as the secure system when the number of attackers is less than 3. However, the system performance is not completely free from degradation. For example, in Figure 5b, when the number of attackers is 2 and 3, the system energy efficiency is much lower than when the number of attackers is 0. And with the increasing number of attackers, the performance of the system may further degrade.

Fig. \ref{fig5-3} and Fig. \ref{fig5-4} show the system throughput and energy efficiency under the data poisoning attack. It can be observed that the data poisoning attack is less disruptive to system performance compared with the model poisoning attack. This is because the data poisoning attack is performed by indirectly changing model parameters through data and the model poisoning attack is directly performed to model parameters. Therefore the model poisoning attack is more aggressive. The result of the data poisoning attack also shows that our proposed choice-based framework is more robust. When the number of attackers is less than 5, it achieves a comparable level of performance as the secure system.

Finally, we further illustrate how choice-based cooperation tuning works in defending against attacks through Fig. \ref{fig6}. During the simulation, we start with four malicious participants and reduce the number of attackers one by one. In this process, we evaluate the system risk level according to the Kullback–Leibler (KL) divergence between the global model output and the local model output. When the KL divergence exceeds the threshold value, the participant will perceive the system as being at risk. The risk assessment results of all participants are averaged as the estimated system risk level. The cooperation level is assessed by measuring the number and duration of participants involved in the FL cooperation. At the beginning of simulation, the number of malicious participants is high. As a result, there is a big difference between the globally shared knowledge and the local knowledge. So the KL divergence is large and it indicates a higher system risk level. In this case, the participants will choose to be less involved in the FL cooperation and their cooperation level will be low. When the number of malicious participants decreases, the system risk level will decrease, and the participants can better benefit from the cooperation. As a result, the cooperation level increases. In this way, participants can work cooperatively while keeping the local model secure during attacks. It is worth mentioning that our proposed approach is a preventive method. It aims to perform the protection when the attacker has the intention to attack rather than after it compromises the system. Hence the attack and the countermeasures happen in the same turn with no delay in between.

\section{Conclusion}

FL is a promising technology for the design of privacy-preserving wireless network applications. In this work, we first make an exhaustive investigation and summarize the under-explored benefits and concerns of applying FL to wireless networks. Next, we generalize the essence of existing personalized FL frameworks and present for the first time the idea that participants have a tunable level of cooperation in personalized FL. On this basis, we propose a novel choice-based FL framework and discuss how tunable cooperation in FL helps solve the key concerns and protect participants from malicious attacks. The case study of knowledge distillation-based cell sleep control further demonstrates the effectiveness of our proposed framework in defending against poisoning attacks and improving the robustness of FL.


\section*{Acknowledgment}
This work has been supported by MITACS and Ericsson
Canada, and NSERC Canada Research Chairs program.

\begin{refcontext}[sorting = none]
\small
\printbibliography

@article{li2022learning,
  title={Learning to attack federated learning: A model-based reinforcement learning attack framework},
  author={Li, Henger and Sun, Xiaolin and Zheng, Zizhan},
  journal={Advances in Neural Information Processing Systems},
  volume={35},
  pages={35007--35020},
  year={2022},
  month={10}
}

@article{tan2022towards,
  title={Towards personalized federated learning},
  author={Tan, Alysa Ziying and Yu, Han and Cui, Lizhen and Yang, Qiang},
  journal={IEEE Transactions on Neural Networks and Learning Systems},
  year={2022},
  month={5},
  publisher={IEEE}
}

@inproceedings{hong2023federated,
  title={Federated robustness propagation: sharing adversarial robustness in heterogeneous federated learning},
  author={Hong, Junyuan and Wang, Haotao and Wang, Zhangyang and Zhou, Jiayu},
  booktitle={Proceedings of the AAAI Conference on Artificial Intelligence},
  volume={37},
  number={7},
  pages={7893--7901},
  year={2023},
month={6}
}

@article{kairouz2021advances,
  title={Advances and open problems in federated learning},
  author={Kairouz, Peter and McMahan, H Brendan and Avent, Brendan and Bellet, Aur{\'e}lien and Bennis, Mehdi and Bhagoji, Arjun Nitin and Bonawitz, Kallista and Charles, Zachary and Cormode, Graham and Cummings, Rachel and others},
  journal={Foundations and Trends{\textregistered} in Machine Learning},
  volume={14},
  number={1--2},
  pages={1--210},
  year={2021},
month={6},
  publisher={Now Publishers, Inc.}
}

@ARTICLE{zhang2023device,
  author={Zhang, Han and Zhou, Hao and Elsayed, Medhat and Bavand, Majid and Gaigalas, Raimundas and Ozcan, Yigit and Erol-Kantarci, Melike},
  journal={IEEE Transactions on Cognitive Communications and Networking}, 
  title={On-Device Intelligence for {5G RAN}: Knowledge Transfer and Federated Learning Enabled {UE}-Centric Traffic Steering}, 
  year={2024},
  volume={10},
  number={2},
  pages={689-705}}

@inproceedings{zhang2023distributed,
  title={Distributed Attacks over Federated Reinforcement Learning-enabled Cell Sleep Control},
  author={Zhang, Han and Zhou, Hao and Elsayed, Medhat and Bavand, Majid and Gaigalas, Raimundas and Ozcan, Yigit and Erol-Kantarci, Melike},
  booktitle={2023 IEEE Globecom Workshops (GC Wkshps)},
  year={2023},
month={12}
}

@article{ghosh2020efficient,
  title={An efficient framework for clustered federated learning},
  author={Ghosh, Avishek and Chung, Jichan and Yin, Dong and Ramchandran, Kannan},
  journal={Advances in Neural Information Processing Systems},
  volume={33},
  pages={19586--19597},
  year={2020},
month={12}
}

@article{arivazhagan2019federated,
  title={Federated learning with personalization layers},
  author={Arivazhagan, Manoj Ghuhan and Aggarwal, Vinay and Singh, Aaditya Kumar and Choudhary, Sunav},
  journal={arXiv preprint arXiv:1912.00818},
  year={2019},
month={12}
}

@article{hanzely2020federated,
  title={Federated learning of a mixture of global and local models},
  author={Hanzely, Filip and Richt{\'a}rik, Peter},
  journal={arXiv preprint arXiv:2002.05516},
  year={2020},
month={2}
}

@article{smith2017federated,
  title={Federated multi-task learning},
  author={Smith, Virginia and Chiang, Chao-Kai and Sanjabi, Maziar and Talwalkar, Ameet S},
  journal={Advances in neural information processing systems},
  volume={30},
  year={2017},
month={12}
}

@article{chen2020fedhealth,
  title={Fedhealth: A federated transfer learning framework for wearable healthcare},
  author={Chen, Yiqiang and Qin, Xin and Wang, Jindong and Yu, Chaohui and Gao, Wen},
  journal={IEEE Intelligent Systems},
  volume={35},
  number={4},
  pages={83--93},
  year={2020},
    month = {7},
  publisher={IEEE}
}

@inproceedings{zhu2021data,
  title={Data-free knowledge distillation for heterogeneous federated learning},
  author={Zhu, Zhuangdi and Hong, Junyuan and Zhou, Jiayu},
  booktitle={International conference on machine learning},
  pages={12878--12889},
  year={2021},
  organization={PMLR},
month={7}
}

@article{hu2021reward,
  title={Reward shaping based federated reinforcement learning},
  author={Hu, Yiqiu and Hua, Yun and Liu, Wenyan and Zhu, Jun},
  journal={IEEE Access},
  volume={9},
  pages={67259--67267},
  year={2021},
  publisher={IEEE},
month={4}
}

@ARTICLE{6056691,
  author={Auer, Gunther and Giannini, Vito and Desset, Claude and Godor, Istvan and Skillermark, Per and Olsson, Magnus and Imran, Muhammad Ali and Sabella, Dario and Gonzalez, Manuel J. and Blume, Oliver and Fehske, Albrecht},
  journal={IEEE Wireless Communications}, 
  title={How much energy is needed to run a wireless network?}, 
  year={2011},
  volume={18},
  number={5},
  pages={40-49},
  doi={10.1109/MWC.2011.6056691}}

@inproceedings{mcmahan2017communication,
  title={Communication-efficient learning of deep networks from decentralized data},
  author={McMahan, Brendan and Moore, Eider and Ramage, Daniel and Hampson, Seth and y Arcas, Blaise Aguera},
  booktitle={Artificial intelligence and statistics},
  pages={1273--1282},
  year={2017},
  organization={PMLR}
}
\end{refcontext}

\end{document}